\documentclass[twocolumn]{aa}  % for a paper on 2 columns
\usepackage{graphicx}
\usepackage{txfonts}
\usepackage{amssymb}
\usepackage{subfigure}
\usepackage{psfrag}
\usepackage{color}

\usepackage{natbib}
\bibpunct{(}{)}{;}{a}{}{,} % to the follow A&A style

\def\Images{.}
\def\drafttype{false}   % Non => Images incluses (Pour sortie finale)

\begin{document}
\title{2D numerical study of the radiation influence on shock structure relevant to laboratory astrophysics}

\author{Matthias Gonz\'alez \inst{1,2,3}, Edouard Audit \inst{2,3} \and Chantal Stehl\'e \inst{4} }

\offprints{M. Gonz\'alez}

\institute{Instituto de Fusi\'on Nuclear, Universidad Polit\'ecnica de Madrid, Madrid, Spain
\and
Service d'Astrophysique, CEA/DSM/IRFU/SAp, Centre de Saclay, F-91191 Gif-sur-Yvette, France
%CEA, IRFU, SAp, Centre de Saclay, F-91191 Gif-sur-Yvette, France
\and
Laboratoire AIM, CNRS, CEA/DSM, Universit\'e Paris Diderot, F-91191 Gif-sur-Yvette, France
\and
LERMA, Observatoire de Paris, Universit\'e Paris VI, CNRS, 5 place J. Janssen 92195 Meudon.
}

\date{Received ???; accepted ???}

\abstract
%context
{Radiative   shocks  are found in  various   astrophysical objects and
particularly at different  stages  of  stellar  evolution.    Studying
radiative shocks,  their  topology, and   thermodynamical properties is
therefore a starting  point to understanding their physical properties.
This study has become possible with the development of large laser
facilities,  which has provided fresh impulse to laboratory
astrophysics.}
%aims
{We present the  main characteristics of radiative shocks
modeled using cylindrical simulations.
We focus our discussion on
the importance of  multi-dimensional radiative-transfer effects on  the
shock topology and dynamics.}
%methods
{We present  results  obtained with our code HERACLES for
conditions corresponding  to experiments   already performed on  laser
installations.  The  multi-dimensional hydrodynamic  code  HERACLES is
specially   adapted  to laboratory   astrophysics  experiments  and to
astrophysical  situations  where  radiation   and   hydrodynamics  are
coupled.}
%results
{The importance of the ratio of the photon mean free path to the
transverse  extension of the shock is emphasized.  We present how it is
possible  to achieve  the  stationary limit of   these  shocks in  the
laboratory  and  analyze the angular   distribution of the radiative
flux that may emerge from the walls of the shock tube.}
%conclusion
{Implications  of    these studies  for   stellar  accretion  shocks are
presented.}


\authorrunning{Gonz\'alez, Audit \& Stehl\'e}
\titlerunning{2D numerical study of radiative shocks}

\maketitle 

%**********************************************************************
\section{Introduction}
%**********************************************************************
Radiative shocks are  shocks in which the  structure of  the flow is
affected by  radiation, either because the radiative energy density
 cannot be neglected  or because the  energy transport  by radiation
(i.e.,   the radiative flux) is significant.   For densities relevant to
laboratory astrophysics,   the  radiative flux becomes   important at
temperature of a few tens of  eV. The temperature  has to be about one
order of magnitude higher before the radiative  energy becomes  important.
In astrophysics and in particular stellar physics, the density can
be far lower and radiation therefore   becomes important  at lower
temperatures.

A characteristic of  radiative shocks is that at high
shock  speeds,  the shocked  matter is  strongly  heated  and  radiates
energy.  Unless the cold upstream gas is completely
transparent,
it absorbs  this radiation creating a hot  radiative
precursor  in front   of the  shock  where   the matter is   generally
ionized.
We focus hereafter on shocks that exhibit a developed radiative precursor.

The  modeling of these radiative shocks is difficult because of the
 nonlocal   coupling  between hydrodynamics   and   radiation,  of the
different  scales      important to  radiation     and   hydrodynamics
\citep{zeldovich, sincell1, sincell2}, and of multi-dimensional effects
\citep{leygnac, lpb}.

The topology  and  evolution of  radiative  shocks depend  on several
factors,
 such as
the  geometry  (from 1D, as  in  shock tubes, to  more
complex  geometries),   the energy  deposition  (impulsive,   as for
radiative blast   waves, or continuously driven   by a  constant piston
velocity),  the optical depth  of the  downstream and upstream regions
relative  to the  shock, and  the possible  contribution of a magnetic
field.  A    detailed  description  of  radiative     shocks is given by
\citet{zeldovich}, \citet{mihalas}, and \citet{drake}.

These strong radiative shocks are found  at different stages of
stellar  evolution: from star  formation, when the accreting gas falls
onto the  protostellar forming object;
during the accretion -  ejection process, which manifests  itself
in terms of the jets of Young Stellar Objects (YSO)  and stellar disks, in
the  atmospheres   of  evolved stars,    when   shocks drive pulsation
\citep{chadid,   fadeyev1,  fadeyev2}; and   in  the  last  stages  of
supernovae explosions.

Spherical  or 1D symmetry  is  the usual obvious approximation,  which
prevails  in the   lack of further  information  about the geometry   of  these
astrophysical  shocks. However, if   this approximation  is reasonable
 as a  first  step  for pulsating   stars, accretion shocks,   and
supernovae explosions, it  may  become  questionable  in the case   of
accretion funnels  of classical  T-Tauri stars,  where the  matter  is
supposed to fall from the stellar disk to  the stellar surface by means of
channels that  follow  the  magnetic-field  lines.  Accretion in
Classical  T-Tauri stars is inferred from  veiling in the spectrum and
also from signatures in the spectral lines.  Simulations indicate that
the topology of  the  accretion funnels  is strongly dependent  on the
relative orientation of the magnetic-field  symmetry and rotation axis
\citep{romanova}.  At the  base  (on the stellar surface),  the column
may vary from compact circular  to ring geometries, whereas the  upper
part (connected to  the disk) may exhibit various  fold-like,  or more
compact topologies driven by instabilities.   This topic is still the
subject of  much debate and controversy.
	
Both the complexity in the structure
of the radiative shocks, even for a 1D approximation, and the lack of angular resolution, require
accurate simulations coupling radiation transport with
hydrodynamics.  This motivates the study of radiative shocks in 2D
simple cylindrical geometries, to analyze the effect of the  finite
lateral size of the radiative shock, on its structure and luminosity.
One should not use identical simulations to both check with and
 analyze  observations.
Experimental   validation of   these  complex hydrodynamic features is thus required.

Obviously, the typical  dimensions of astrophysical radiative  shocks
ensure that they cannot be studied directly  in the   laboratory.  The study  of
radiative  shocks on Earth  thus  require different gases and physical
conditions, which can be recreated by   high energy installations  such as
laser   or  pulsed electric   installations  \citep{remington}.
Radiative shock  experiments  have  been  performed  with high power
lasers  \citep{bozier1,  bozier2,   keiter,  reighard, bouquet,   lpb,
busquet, busquet2}.  Typically, a 200~J  laser in about 1~ns can
launch a shock at about 60~km/s  in targets  of  millimeter
size \citep{bouquet} filled with xenon  at pressures of some fractions
of  bars,   whereas supercritical  shocks   at 100~km/s   in SiO$_{2}$
aerogel, argon,   and xenon have   been  produced  at  the OMEGA laser
(5~kJ).  Strong shocks  have also been generated by a compact
pulse-power device \citep{kondo}, producing  shocks at 45~km/s in xenon
with different gas pressures of up to  10$^{-2}$~bars. We see below
that, for   a  fixed  shock  velocity  accessible  to a laser
installation, the radiative regime is more easily  achieved with a
low-density, high-atomic-mass gas, which explains why low-density xenon is
often used in these studies.

These experiments   enable one to   examine  the  physics of  the
radiative precursor (dynamics  and density) by visible  interferometry
and provide information about the temperature  of the shock front, by
visible emissivity.  Studies of the dynamics of the precursor of xenon
radiative shocks  by interferometry \citep{bouquet}, and shadowgraphy
\citep{lpb}, and of its  topology by instantaneous X-ray imaging
\citep{vinci} indicate  that multi-dimensional effects can affect the shock
wave, and, in  particular,  its precursor.  This  was  attributed to the
lateral  radiation losses (through   the  walls of the  shock tube),
which reduce the amount of radiation heating  the precursor and thus affect its
structure \citep{leygnac, lpb}.

\citet{keilty}, \citet{shigemori}, \citet{edwards}, \citet{calder}, and \citet{laming} focused   their studies
on the   spherical,    radiative,  blast  waves,   which  are similar 
to those studied in the previous 1D    cases, in terms of the development   of
radiative precursor, but     also exhibit strong  differences, such as      the
importance    of  radiative cooling  and   the  development of various
instabilities.  The study of these  experimental blast waves is linked
to the  problem   of  the  evolution and stability    of supernova
remnants.

After the   presentation  of the  typical   structure  of 1D radiative
shocks,   we present a numerical  study of laboratory
radiative  shocks  that can    be   generated by medium-power   laser
facilities, such  as  PALS (Prague, Czech   Republic), LULI (Palaiseau,
France), and similar lasers.   The numerical code (HERACLES) used
in this work is first briefly presented.  We then study the effect of
the   duration    of   the  shock   driven   by the    piston
(Sect.  \ref{sec:imp}).  In   Sect.~\ref{sec:width},  we  consider   the
effects   of  multi-dimensional  radiative  transfer   and their
correlation with the width of the shock tube.
In Sect.~\ref{sec:albedo}, we
examine the influence of the  radiation boundary-conditions,
which are  inferred from the  albedo of the cell wall,  on the
topology  and   propagation of the  shock.    Finally,  we  present the
perspectives  and   conditions  for obtaining  an  experimental  stationary
radiative  shock and present the   angular dependence of the  emerging
radiative flux of such a shock.

%**********************************************************************
\section{Structure of a 1D radiative shock}
%**********************************************************************
% simu sur dapax40 /tmp2/mgonzale/PALS_EOS/1D_NEW

In this section, we present the typical features  of a radiative shock
in terms of 1D geometry. When  the flow passes through  the shock, it is heated
and it re-emits a fraction of its energy  by radiation processes. These
emitted photons are  then absorbed  by the  cold and opaque  unshocked
gas.  This    preheated    upstream gas  is     called   the radiative
precursor. When the stationary regime is reached, a radiative shock is
then composed of  both a purely hydrodynamical  shock (which consists  of a
discontinuity in the  hydrodynamical quantities)  and  a  radiative
precursor, whose length is directly  correlated with both the mean free path
of photons in  the gas and  the shock velocity.   One usually
considers two classes of shocks  depending on the temperature  reached
just  ahead of the hydrodynamical discontinuity.  If it is lower than the
post-shock temperature, the shock is called  subcritical, and if it is
equal the shock is called supercritical.

By assuming that a strong shock develops in an initially cold
gas  of    density  $\rho_0$  (mass number $A$)    and  solving the
Rankine-Hugoniot relations, one  can show that the temperature reached
in the shock equals 
\begin{eqnarray}
\begin{array}{ll}
T = 120.27  \; {\rm K} \;  \frac{A}{<Z>+1} & \left(\frac{u_s}{1 \, {\rm km/s}}\right)^2  \frac{(\gamma-1)}{(\gamma+1)^2} \\
& \times \left[1 + \sqrt{1+(\gamma^2-1)X} -(\gamma+1)X  \right], \\
\end{array}
\label{eq:tchoc}
\end{eqnarray}
where $<Z>$ is the mean ionization stage in the shocked material, 
$X  = \frac{e_i}{\frac12 \mu u_s^2}$   is the ratio of the  ionization
energy per  atom to the initial  kinetic energy per atom,  and $\mu=$
the mean molecular weight of the non-ionized gas).
The photons  escaping from the shock front  are used to
heat  the cold gas, incoming onto  the front, to a temperature of $T_{-}$
(which is  lower than  $T$).  Neglecting the   gas compression in  the
precursor, one has \citep{zeldovich}:
\begin{eqnarray}
\sigma T^4 = \rho_0 u_s \epsilon (T_{-}, \rho_0),
\label{eq:t4}
\end{eqnarray}
where  $\epsilon=\frac{kT(1+Z)}{\mu(\gamma-1)}+\frac{e_i}{\mu}$ is the internal specific energy  (erg/g)  after taking into account the ionization. 

When $  T_{-}$ equals $T$, the  shock enters the  supercritical regime,
which is characterized by  an  extended precursor.  The  corresponding
critical velocity is obtained formally by resolving the coupled set of
Eqs.~(\ref{eq:tchoc})  and (\ref{eq:t4}).  This system must be solved
numerically  because one needs  to take account of the
temperature  dependence of both the internal  energy and ionic charge.
Using  the atomic-physics model described   by \citet{michaut} and for
xenon  of density  10$^{-3}$  g/cm$^3$, we found  a  critical  velocity around
30~km/s, achieving  a shock  temperature  of 12~eV  and  average ionic
charge of 8.   In the particular  case  of a perfect  gas with $\gamma
=5/3$, and neglecting the ionization-energy term in Eqs.~(\ref{eq:tchoc})  and (\ref{eq:t4}), one obtains the
following approximate value of
\begin{eqnarray}
v_s \ge 450 \; {\rm km/s} \; \left(\frac{<Z>+1}{A}\right)^{4/5}  \left(\frac{\rho_0}{1 \, {\rm g/cm^3}}\right)^{1/5}
\label{eq:vs}
\end{eqnarray}
which illustrates that the supercritical regime is more easily obtained at
low  density  and  for  high-mass  gases. Using  1D hydro-radiative
simulations, \citet{ensman}  found  a supercritical  regime at 20~km/s
for  hydrogen at  8x10$^{-10}$~g/cm$^3$ in  qualitative agreement with
Eq.~\ref{eq:vs}. 

The  precursor  extension, $L$,  can be
approximated with a radiative heat-conduction model.  One then finds that
$L$ is  about several photon  mean-free-paths close to the
supercritical limit \citep{nemchinov, penzo}.

\begin{figure}
\includegraphics[height=0.97\hsize,draft=\drafttype,angle=90] {\Images/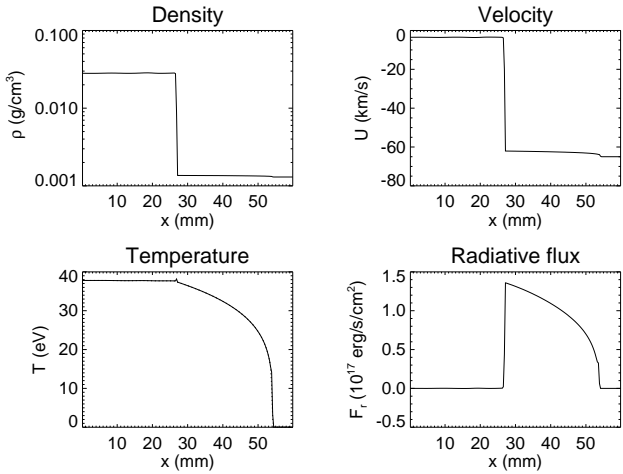}
\caption{\label{fig:choc1D} Typical profiles of a 1D radiative shock: 
density  (upper left),    velocity  (upper right),  radiative    and gas
temperature (lower left), and radiative flux (lower right).}
\end{figure}

A    typical  1D   supercritical    shock   structure   is shown    in
Fig.~\ref{fig:choc1D}.   These academic  profiles were  obtained using
xenon with   a constant ionisation   charge  of $5$, a  perfect-gas
equation-of-state   with   $\gamma$=1.1,   and    analytic opacities
\citep{bozier1}.    The velocity of  the shock   is  65~km/s and the
initial density and    temperature are   1.3x10$^{-3}$~g/cm$^3$ and
300~K, respectively. The radiative  precursor is clearly evident in the
radiative flux  and   temperature profiles but,  at this moderate  shock
speed, the density and  velocity are only slightly affected.

%**********************************************************************
\section{The HERACLES code} \label{sec:heracles}
%**********************************************************************

The  simulations  presented  in  this   paper  were    completed using  the
three-dimensional     radiation-hydrodynamics     code        HERACLES
\citep{heracles}.  This code   solves  the  Euler equations  for 
hydrodynamics  coupled with  the  moment    equations of  the   transfer
equation. The  original closure relation  used for the radiative model
(M$_1$ model) allows  one to  be exact both  in the diffusive and
 transport  limit regimes.    
This code output data was verified by analytic models and compared with
other codes.
It can
study     a     wide   variety     of      astrophysical    problems
\citep{heracles}. Furthermore,  it was  used to   develop and
analyze a laboratory experiment  of radiative shocks \citep{lpb}. This
cross-validation between experiment and simulation illustrates the
 relevance of  HERACLES  to  both  laboratory astrophysics and   
classical astrophysical situations.

The strength  of HERACLES is that it  can deal  with multi-dimensional
problems.  Until now, radiative  shocks have been studied  by 1D
geometry models,  multi-D models with hydrostatic  flows, or by adopting the flux-limited
diffusion   approximation (\citealp{bouquet}, \citealp{drake2007}  and
reference  therein).  On the  one hand,  the multi-D effects
(such  as lateral  losses) can    determine the structure  of the
flow,  and   on the  other  hand, at  the foot  of  the radiative
precursor,  the reduced flux  (ratio of  the radiative  flux  to the product of the
radiative energy and the light  speed) is nearly one so that
we are  in the transport limit.   In this  context, HERACLES is
of particular relevance to the study of multi-D radiative shocks.

Although all    shocks in  reality propagate  within 3D media,  we 
performed 2D axisymmetric simulations. As  a  first   step, we considered only
the   determination of     the theoretical
influence of   different  parameters  on   the shock propagation.
Furthermore,     we    verified that  2D   axisymmetric and 3D
Cartesian simulations produced identical
results (e.g., speeds, positions)  if one   is careful to ensure that
the  surface-to-volume  ratio is equal in both   cases .  Our  results  are
therefore relevant  to the case of radiative  shocks propagating in a
rectangular medium, as in most laboratory experiments.

Hereafter, all  simulations presented    are
for shocks propagating in cylindrical cells with the left vertical
boundary corresponding   to the symmetry  axis and  the lateral losses
located at the right vertical boundary.
They  are filled with xenon,  which is initially at ambient temperature
and  a pressure of  0.2~bar (i.e., this  corresponds to a density of
1.3x10$^{-3}$~g/cm$^3$).  These   values  are   typical  of laboratory
experiments \citep{bouquet, lpb} where  a high molecular weight and  a
low density are chosen  to maximize radiative effects.  If
not    specified,  we use a  realistic   equation  of state  and
opacities.  The  equation  of state  of xenon was  computed using the
OPA-CS code of C.  Stehl\'e, which uses a screened hydrogenic model and
is described in \citet{michaut}.
 The opacities were kindly provided by M. Busquet, who computed them using the STA code \citep{barshalom}.
The  shock is driven at  a speed of 65~km/s which  is  a typical
 of present-day  laser  experiments.  
The  boundary of the  cells can have  a variable albedo, defined as
the fraction of  the incident radiative  flux re-emitted by
the   wall.  We demonstrated in  previous work \citep{lpb}  that this
parameter has a strong influence on the precursor propagation. We 
therefore performed simulations   for    albedos between  0$\%$
(fully transparent or fully absorbing boundaries) and 100$\%$ (total
 reemission).

%**********************************************************************
\section{Launching effect} \label{sec:imp}
%**********************************************************************
%simu sur dapax40  /tmp2/mgonzale/PALS_EOS/SANSEOS_AVECFUITES_INJcont/

The  dynamics   of the  shock   propagation  depend  strongly  on the
launching  procedure. In laser driven shocks, the conversion of the laser energy to the piston mechanical energy is performed by the shock generated in the piston during the laser ablation \citep{bouquet, drake}.
In this paragraph   we compare the dynamics of the shock for two launching scenarios.
In the first, the piston moves at a constant velocity.
The second corresponds to the case where the piston decelerates with time.
Hereafter, we denote these two driving cases as continuous and impulsive respectively.

\begin{figure}
\centering
\begin{minipage}[c]{0.5\hsize}
\centering 
\subfigure[Shock and precursor absolute positions]{
\includegraphics [height=\hsize,draft=\drafttype,angle=90]
		 {\Images/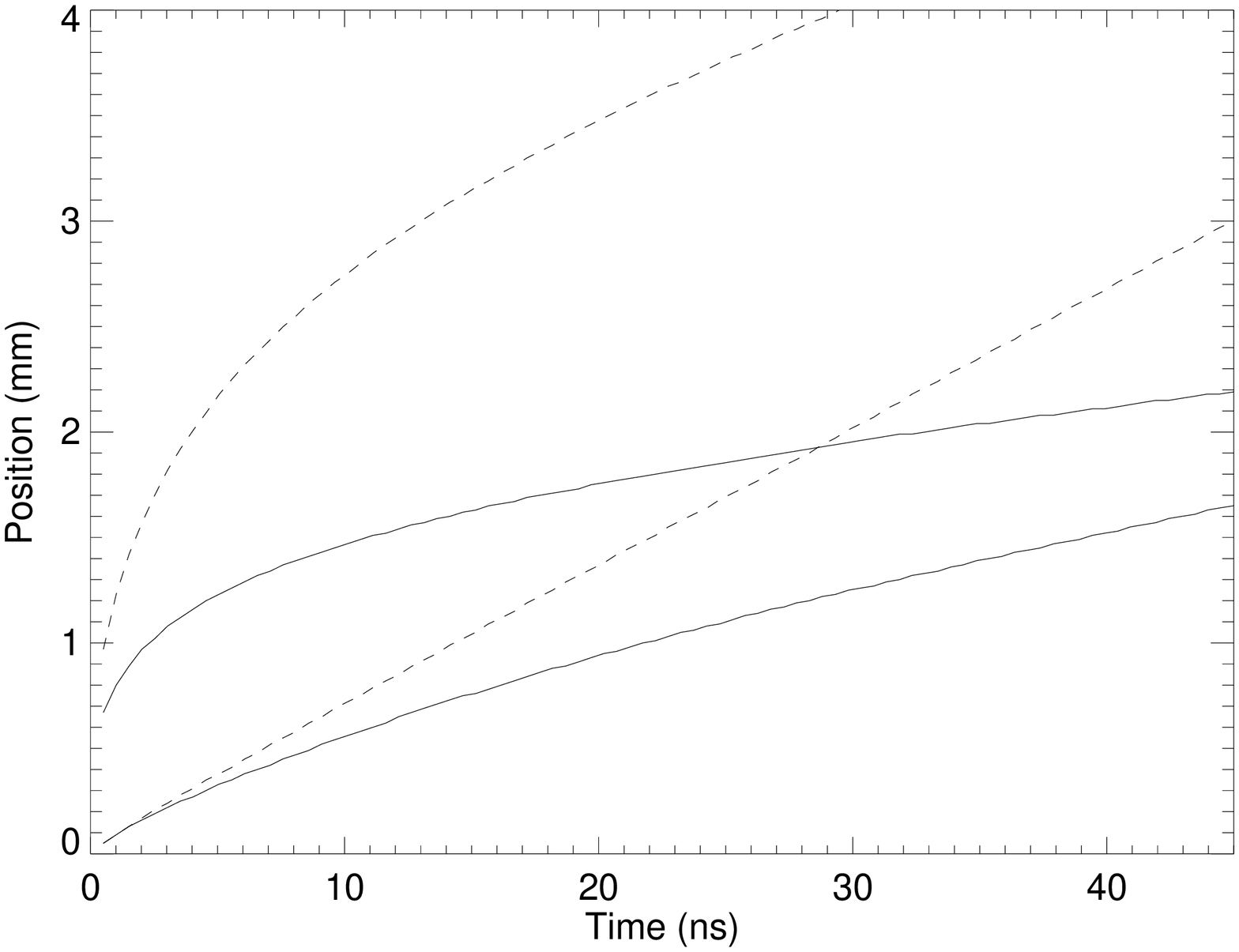}
\label{fig:choc_contcoup1}}
\end{minipage}%
\begin{minipage}[c]{0.5\hsize}
\centering 
\subfigure[Distance between shock and precursor]{
\includegraphics [height=\hsize,draft=\drafttype,angle=90]
		 {\Images/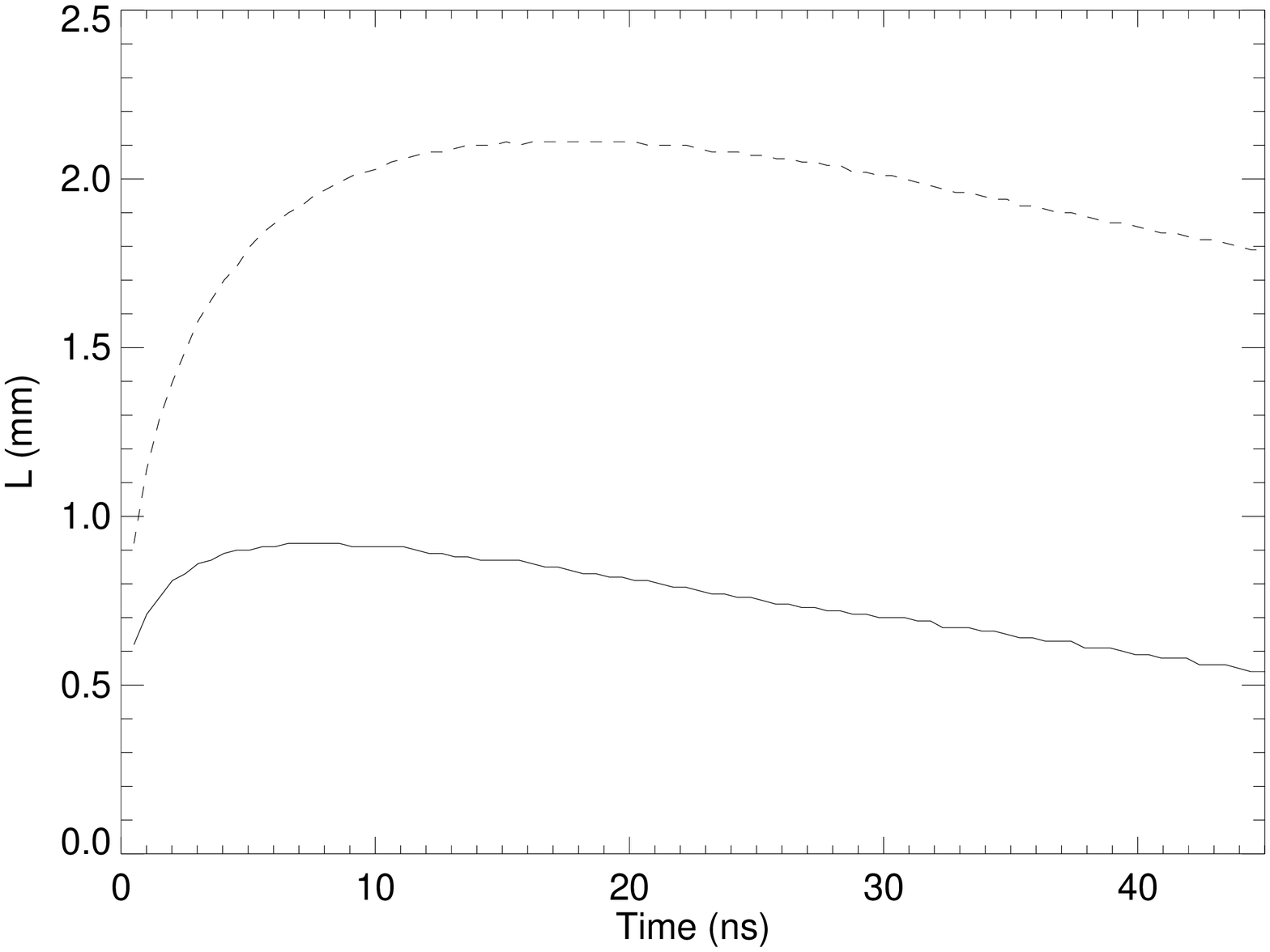}
		 \label{fig:choc_contcoup2}}
\end{minipage}
\caption{Influence  of  the dynamics   of the  piston  launching  on the  shock
dynamics.   The  shock  is  driven in   xenon   at 0.2~bar    at room
temperature. The  shock  tube is a  cylindrical  cell  with a 0.7~cm diameter and
albedo of 40\%. The full line refers to the impulsive lanching whereas
the dotted line  refers  to a shock   driven by a piston  at  constant
velocity    of    65~km/s     (cf.   Sects.~\ref{sec:heracles}   and
\ref{sec:imp}).}

\label{fig:choc_contcoup}
\end{figure}

Figure~\ref{fig:choc_contcoup1}  shows the difference  between a shock
with a launching  phase of the shock that is impulsive  and one for which the driving   is continuous.
In  both  cases,  the  lateral 
albedo is 40\%.  When the  shock driver (``piston") is  continuous, the shock
propagates at  constant  velocity equal to the initial velocity of
65~km/s. 
In the impulsive case, the driver of 65~km/s is continuous during 0.3~ns and then
decreases with a temporal dependence that reproduces qualitatively the velocities of the experimental shocks \citep{lpb}.
In that case,
 the  shock decelerates as soon
as the  drive  stops because it has  little  inertia. Regardless of  the drive duration, one    can
observe three different stages  in the
precursor dynamics.  In the first stage, the precursor moves faster than
the shock. The photons emitted by the shock  propagate in the upstream
fluid, which is optically thick, where they are  absorbed.  This  induces a
warm, ionized, radiative precursor in the upstream part of the shock. In
a second phase, the precursor decelerates (it can be even slower than
the shock front). Finally,  it reaches a constant  speed equal to that
of the shock front.  This is the stationary regime.

These     three     stages     are     clearly     seen     in     
Fig.~\ref{fig:choc_contcoup2} which shows  the distance separating the
shock from  the precursor leading-edge versus time. Since  the precursor  goes faster
than the shock,  this distance increases  to  a maximum value. Then,
as the  precursor slows down, this distance  decreases until reaching an
asymptotic value (stationary limit), when both the shock and precursor leading-edge travel
 at  the same speed.

In a  laboratory  experiment, the duration   of the  driving is  not a
parameter  that one can  vary simply.
It  is  easier to modify the  shock  dynamics by changing the
radiative lateral losses, which  can be achieved by varying the width of the
canal or by modifying the albedo of the  cells walls.  This albedo can
be specially adapted    by choosing properly  the  material  of the   cell or by
choosing special coating on these walls.

In  the  following  sections, we    examine   the  influence  of these
parameters on the dynamics of the shock precursor and in particular on
the delay before the stationary regime.
 
%**********************************************************************
\section{ Effect of  a 2D radiation field } \label{sec:width}
%**********************************************************************
We define R to be the ratio of the transverse width l to the
photon mean-free-path $\lambda$ in the unshocked
medium ($\lambda   \simeq 1$~mm).  In this  section, we   present  the
influence of  R on the radiative shock  structure and evolution.
In our bi-dimensional and  axisymmetric simulations, the gas is driven
at a speed of 35~km/s,  constant both in time and across the cell. Since
we impose  lateral reflecting boundaries on the hydrodynamics, the shock
would remain planar  without  accounting for the  radiative transfer.
It would be transversally  homogeneous,  and temperature and   density
would vary only with y-position (along the canal axis).  However, when
accounting   for   radiative  transfer with  lateral    losses through
partially   reflective walls,   2D effects  can be observed.  Density  and
temperature vary with x-position (perpendicular   to the axis) and,  in
some conditions, curvature of the shock front may even be
observed.
The influence of the radiation on the shock structure is directly correlated with
 the possibility of photon escape from the medium, which can be
estimated from the dimensionless number R=l/$\lambda$. 

The   importance of  radiative losses   on  the shock  structure, which
depends strongly on  R, has not been considered  in detail by  many
    laboratory  experiments nor  in the   study of radiative
shocks encountered in astrophysical  objects.   For example, the  mean
free path in xenon at 300~K and $10^{-3}$~g/cm$^3$, which is typical of
laboratory experiments, is  about 0.1~mm whereas the canal width
is about 1~mm.  For a typical stellar atmosphere (T=5000~K
and $\rho$=5x10$^{-11}$~g/cm$^3$), the typical mean free path is
4x10$^{11}$~cm and decreases rapidly when the temperature increases
\citep{seaton} whereas the radius of a classical T Tauri star is about
2x10$^{11}$~cm.

\begin{figure}
\centering
\subfigure[R=0.5]{
\includegraphics [width=0.25\vsize,draft=\drafttype,angle=90]
 		 {\Images/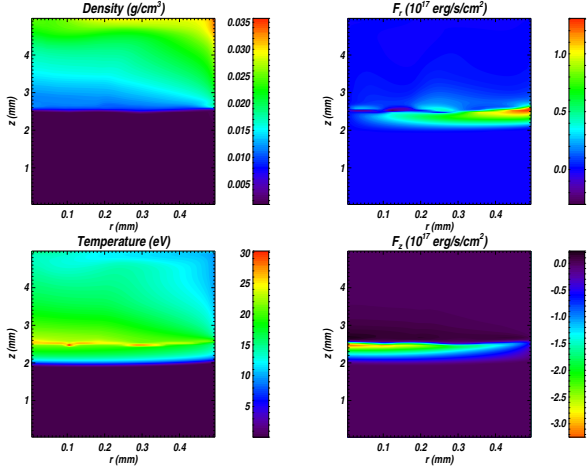}
		 \label{fig:05lpm}}
\subfigure[R=1]{
\includegraphics [width=0.25\vsize,draft=\drafttype,angle=90]
		 {\Images/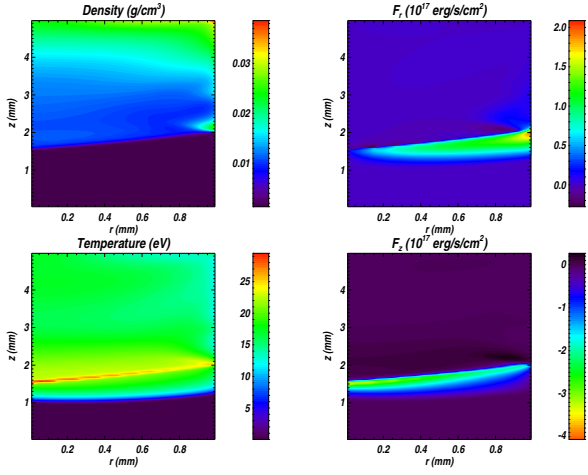}
		 \label{fig:1lpm}}
\subfigure[R=10]{
\includegraphics [width=0.25\vsize,draft=\drafttype,angle=90]
		 {\Images/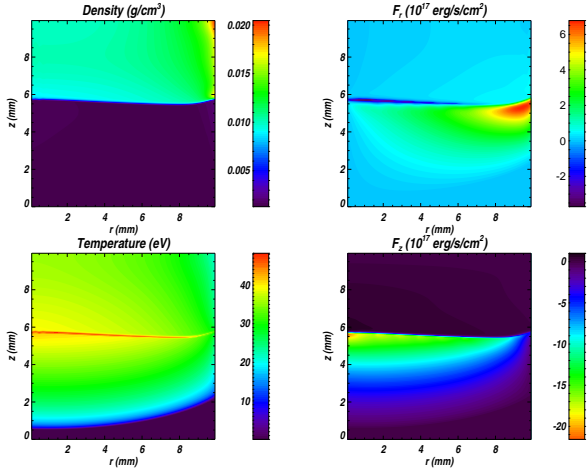}
		 \label{fig:10lpm}}
\caption{At a same instant, snapshots of the maps of density,
temperature, radial, and axial   
radiative flux showing the influence  of  the canal width compared  to
the photon mean free path (R=0.5 top  panel, R=1 middle panel and R=10
bottom panel).  Case of  xenon initially at 1.3x10$^{-3}$~g/cm$^3$
and 300~K, using perfect-gas equation-of-state
($\gamma$=5/3 and mean molecular weight $\mu$=20 equivalent to a mean ionization stage of 5),
and analytic opacities  \citep{bozier1}. The shock propagates from top
to bottom.}
\label{fig:lpm}
\end{figure}

Figure~\ref{fig:lpm} shows bi-dimensional   maps at the  same time for
three different canal  widths   (i.e., three  values  of R).   The
shock position is  marked by the jump  in density
and  by the temperature  peak.  The precursor is  located in the lower
part of the  figure below the  shock.   Its extension  is particularly
visible in the maps of temperature and radiative flux.

If R is small (cf.  Fig.~\ref{fig:05lpm}),  the photons don't interact
with the gas and they escape freely from the system.  As the radiation
escapes laterally,   the shock loses   some  part  of its  energy,  it
decelerates  rapidly, and the precursor  remains thin.  Since the photons
emitted at any point of the downstream medium escape freely, the shock
remains   planar.       In contrast,   if     R    is large  (cf.
Fig.~\ref{fig:10lpm}),  the  photons  cannot escape  because they  are
reabsorbed over a  short distance. The shock  travels faster than in
the first case.  In  the limit where R tends  to infinity,  we recover
the uni-dimensional case with  a planar shock except  for a small
layer  along  the walls,    similar  to a    boundary layer.   In  the
intermediate case (cf. Fig.~\ref{fig:1lpm}),  the shock is curved,  since
only photons  close to the walls   escape, and photons emitted  on the
axis are trapped.  The shock then decelerates more close to the walls than
on  the  axis, which tends  to bend  the shock.   Shock curvatures have 
already been observed experimentally by imaging techniques
\citep{vinci,reighard}, although the dynamical origin of the curvature
can be of many different natures.

%**********************************************************************
\section{Effect of radiative losses} \label{sec:albedo}
%**********************************************************************
In the previous section, we showed that the  ratio R of the photon
mean  free   path to the   canal  width was  a key   parameter for the
radiative    losses.  In this  section, we   study  in more detail the
influence of the  lateral radiative losses on  the shock and precursor
dynamics. We therefore consider a canal of  fixed width but with walls
of variable   albedo  a,   which varies   between  0 (no  wall
reflexion) and 1 (purely reemitting walls).

This wall albedo  is a crucial  point in the  laboratory experiments.
Varying the element used for the cells or covering the cell walls with
layers  of different elements  (e.g., gold,  aluminium), one can test the
radiative-loss effects.

We  study  the influence of  the
albedo on  the  dynamical properties of  the  shock at  short and long
times. To illustrate the effect, we  have chosen to focus the analysis
on the case of  impulsive launching. We concentrate our study
on the  positions of the shock front  and of the  leading  edge of the
precursor  (defined  as the  point, on the cell symmetry axis, where the  temperature reaches the
threshold value of 5~eV). The value of
the temperature threshold influences the extension of
the precursor only slightly.

\begin{figure}
\centering
\subfigure[]{
\includegraphics[width=0.97\hsize,draft=\drafttype]
		{\Images/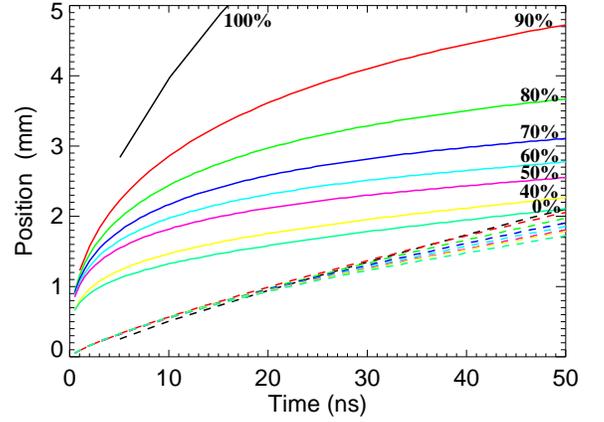}
}
\subfigure[]{
\includegraphics[width=0.97\hsize,draft=\drafttype]
		{\Images/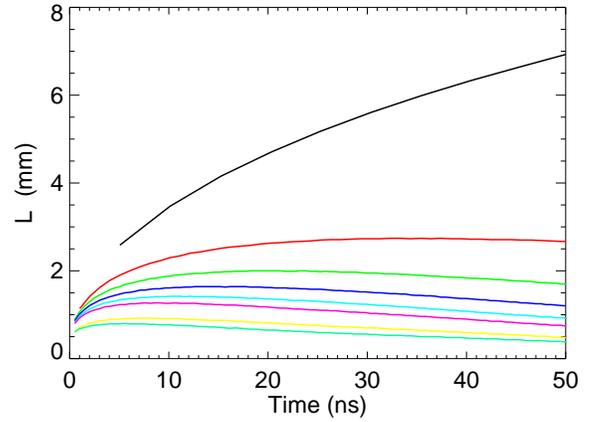}
		\label{fig:choc_albedo_b}
}
\caption{\label{fig:choc_albedo} (a) : Influence of  the walls  albedo upon
the  shock (dashed line) and  precursor  (solid  line) positions.  (b)
Influence  of the albedo  upon the distance  between the precursor and
the shock.  Comparison for an albedo of
 0\%, 40\%, 50\%, 60\%, 70\%, 80\%, 90\%, and 100\% (from bottom to 
top).}
\end{figure}

Figure~\ref{fig:choc_albedo} shows the shock  dynamics obtained with an albedo of 0\%, 10\%, 
20\%, 30\%, 40\%, 50\%, 60\%, 70\%, 80\%, 90\%, and 100\%.
One can see that the  wall albedo has almost no  effect on the shock
velocity,  at least for the evolution  times considered here.  However, the
precursor  is strongly  affected by the   albedo.   As can be seen  in
Fig.~\ref{fig:choc_albedo_b},   for    small values   of    the albedo
(for important lateral  losses),   when most of  the   photons escape, the
precursor  extension  remains     small and its    deceleration 
increases.       This dependence   is     non-linear   (cf.   
Sect.~\ref{sec:stat}  below for  the dependence  of the  time needed to reach the stationary limit
on the albedo).    The extension of the  precursor relative to the
shock   front reduces slowly   with  time,  as   a consequence of  the
weakening of  the shock and the slow  deceleration of the piston. This
evolution at later  times differs from  the case where the piston  
moves  at     a  constant   velocity,      as  can  be    seen   from
Fig.~\ref{fig:choc_contcoup}, which  converges at long times towards a
stationary  limit, where  precursor and  shock front  move at  the same
velocity.  At early times, the precursor front moves far more rapidly
than the shock  front, and we  expect that this regime  is close to the
case of  a  Marshak  wave (a   radiation-driven thermal   wave without
coupling to hydrodynamics).  For t below  5~ns, and a=1, one
thus recovers the   well-known 1D-Marshak-wave time-dependence  of
$\sqrt{t}$.  For smaller values of the albedo, the structure of the 2D-bent
Marshak waves is more complex, even supposing a constant opacity,
as one can see from
\citet{Hurricane}, who studied the bending and slowing  of the front by
radiative losses.  However, even if  similarities exist between the
dynamics and topologies of
a Marshak wave
and those of a radiative   precursor at early  times,  the  two processes
differ physically, especially at  late  times where the stationary
limit is reached in one process and not in the other.

%**********************************************************************
\section{Toward stationary shock} \label{sec:stat}
%**********************************************************************
In the case  of a  shock  with continuous driving,  the typical
evolution of the radiative  shocks exhibits, as in  the previous
case,   a first phase  of  development  of  the  precursor, where the
ionization-front  velocity  exceeds the  shock  velocity.   After this
phase, the ionization-front velocity decelerates until it becomes equal
to the front velocity.  At this time, the stationary limit is reached.
The corresponding time  at which this stationary radiative shock is formed,
$t_{stat}$, is of the order of the ratio of the precursor extension to
the shock velocity.

The stationary shock limit is  usually assumed for astrophysical flows,
because  $t_{stat}$  is    less than  the    typical
hydrodynamical  time, $t_{hyd}$.     For  instance, in  the  case  of
pulsating stars, $t_{hyd}$, which is roughly equal to the ratio of the
atmosphere width to  the shock velocity,  varies between 1500~s for RR
Lyrae and 10$^6$~s for cooler giants such as RV Tauri, whereas $t_{stat}$
is of  the   order  of  50~s   for a  shock  velocity around   10~km/s
\citep{fokin}.  In accretion shocks along magnetospheric columns, this
time  is approximately  equal to  the ratio of the column  length  to the shock
velocity, or  10$^4$-10$^5$ s  for classical T   Tauri stars  (of about 3
solar radius and 300~km/s).  Thus, in  the absence of any shorter
timescale   of    hydrodynamical    motion   (e.g., instabilities),   these
astrophysical shock waves can be   supposed to be stationary from  the
radiative point of view.
 
Associated with the  1D  approximation, this stationary   limit also offers
the  advantage of simplicity in the observational diagnostics
with the help   of  published stationary-shock   structure tabulations
\citep{fadeyev2}.

In the  experimental cases, the  piston velocity decreases slowly
with  time. However,   as in the  academic case   of  continuously
sustained shock, the precursor  extension relative to the  shock front
increases at early   times until reaching   a maximum  value. We  have
chosen to define   the time interval to  form a stationary  shock,
$t_{stat}$, as the time of maximum extension of the precursor because
it unambiguously delimits the short accelerating and long decelerating
phases of the precursor dynamics.

\begin{figure}
\centering
\includegraphics[width=0.97\hsize,draft=\drafttype]
		{\Images/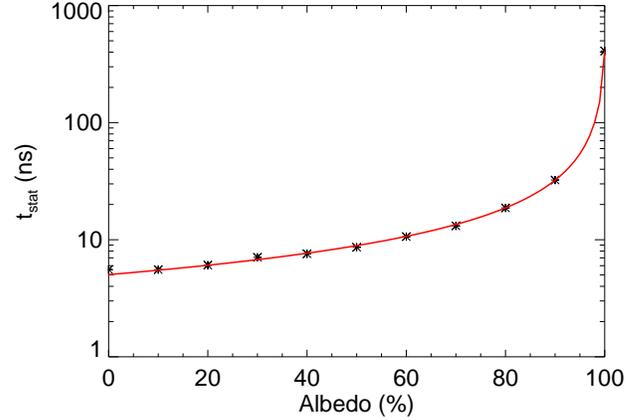}
\caption{\label{fig:stat2} Influence of the albedo upon the time interval to form a stationary radiative shock, $t_{stat}$ (points: numerical simulations results, line: fit).
}
\end{figure}

Using this definition, and in the launching case, which is close to that of
previous experiments \citep{lpb}, we find
$t_{stat}=5.6$~ns, 6.1~ns, 7.1~ns, 7.6~ns, 8.6~ns, 10.6~ns, 13.1~ns, 18.7~ns, 32.3~ns, and 409~ns for wall albedos of 10, 20, 30, 40, 50, 60, 70, 80, 90, and 100 percent, respectively
(cf.~Fig.~\ref{fig:stat2}).
Thus, high lateral radiation losses (small albedo) not only decelerate the radiative precursor, but also strongly decrease the time to reach the stationary limit. From these numerical simulations, we derive an analytical fit 
\begin{equation}
t_{stat}  = \frac{t_0 t_1}{(t_1-t_0)(1-a)^\beta+t_0}
\end{equation}
where $a$ is the wall albedo, $t_0$ and $t_1$ are the stationary times
for a=0 and a=1, respectively, and $\beta$ is a constant. Their values
are $t_0=5.6 \; {\rm ns}$, $t_1= 409 \; {\rm ns}$, and $\beta=0.84$.

In all of these simulations,  the shock position  is insensitive to the
albedo  of the  tube walls.   However, even if the
shock  speed is  constant, the   density  and   temperature of  the
post-shock matter are sensitive to the radiative  losses. Higher
losses  (smaller  albedo)  result in a   higher   density and a  lower
temperature.   As a consequence,  the precursor extension  tends to be
lower in the case of higher losses  because the emissivity of the shock
front is  proportional to $T^4$ and also  because a large  fraction of
the photons escapes without contributing to the precursor extension.

\begin{figure}
\centering
\includegraphics[width=0.97\hsize,draft=\drafttype]
		{\Images/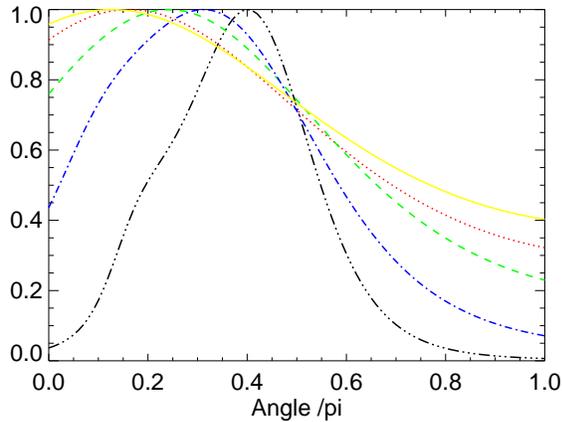}
\caption{\label{fig:flux_lateral3} Normalized luminosities for different values of albedo
(80\% in yellow solid line, 70\% in red dotted line, 50\% in green dashed line, 20\% in blue dashed-dotted line and 0\% in black dashed-double-dotted line).}
\end{figure}

\begin{figure}
\centering
\includegraphics[width=0.97\hsize,draft=\drafttype]
		{\Images/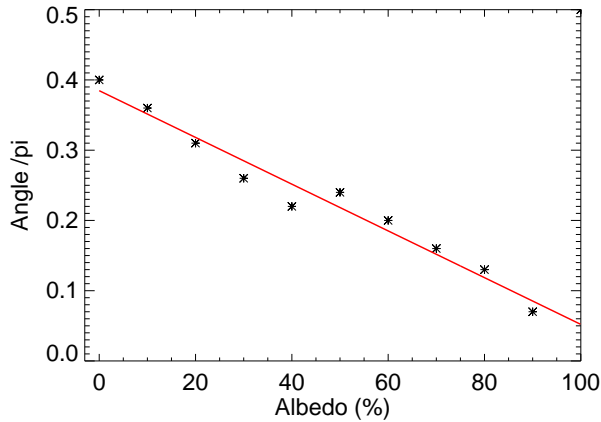}
\caption{\label{fig:flux_lateral2} Angle corresponding to the maximum of the luminosity 
as a function of the albedo
(points: numerical simulation results, line: fit).}
\end{figure}

Until now, we have emphasized the strong dependence of the stationary-shock
structure on its  radiative   parameters  (e.g., ratio R of    the
transverse  width to the photon  mean free path,  and the percentage  of
radiative losses at the  walls of the  shock tube), or equivalently to
departures     from the 1D  approximation.   Another interesting
quantity is the  radiation  flux,  since it  is  used, for instance, as  a
tracer of  accretion rate.   In Fig.~\ref{fig:flux_lateral3}, we  plot,
for different tube-wall albedos,
the  variations  in the  luminosity  as  a  function  of the angle  of
observation    $\theta$  ($\theta=0$ corresponds    to a forward flux,
$\theta=\pi/2$ to a perpendicular  flux, and $\theta=\pi$ to a backward
flux).  The  plotted profiles correspond  to the luminosity integrated
from the beginning of the radiative precursor  to the shock front 
at  time t=50ns. The integral is evaluated by calculating the photon angular
distribution function, which is known from the M1 model, at each point along
the cell wall. We plot  a normalized luminosity
because we are interested only in the  angular distribution and not in
the absolute value, which  is  more difficult to compare. One can
see that with increasing lateral losses the angle corresponding to the
maximum of the  luminosity increases and  the distribution sharpens
around this  angle.  This is  due to the fact  that when the losses
are high, the radiative reduced  flux is higher, and the  luminosity
tends to become  almost mono-directional (in the  limiting case of a
unit-reduced flux) rather than
an isotropic Planck function (in the  limiting case of diffusion). The
luminosity  is much  more directional and  the   dispersion around the
preferred direction   is lower.  Figure~\ref{fig:flux_lateral2} shows
that this angle is approximately a linear  function of the lateral albedo of the tube walls.
All these points  illustrate the strong anisotropy of  the flux in the
radiative shock.
Although illustrated for a given time, the conclusions about the influence of the albedo on the peak of the angular distribution and on the dispersion about this value are of a quite general character during the shock propagation.

%**********************************************************************
\section{Conclusion}
%**********************************************************************

We have   illustrated how  radiation influences the
radiative-shock topology and structure. These effects can be
observed with experimental      results and used  to   interpret
observational data.  

We have  in particular shown  that lateral  radiative  losses induce a
curvature in the shock front and a strong  shortening of the radiative
precursor.   For laboratory  experiments,    it means that the   tube
geometry and the opacities  play a key  role in the shock  propagation
and  topology.  In astrophysical  cases, this   suggests  that the  link
between accretion rate  and photometric  observation is  strongly
dependent on the gas conditions and  the possible presence of dust.

Lateral radiative losses  through  the walls of the tube  lead  to a
weakening of  the shock, and a  reduction in  the precursor extension  and
temperature.  They also strongly reduce  the time needed to reach  the
stationary limit.

The   radiative flux that emerges  from  the  tube boundaries evolves
from an isotropic situation in the case of low radiation losses to a
more anisotropic   distribution  in  the  case  of   strong  radiative
losses. This  effect  may have a  strong impact  on photometric
observations of anisotropic accretion  flows, such as in the
case of T Tauri star accretion through funnels.

High-energy,  high-power  installations,  such as   lasers, allow
generation of strong radiative shocks  with extended precursors.  Most
studies have  focused on the global characterization of the
radiative precursor.  Only a few have attempted to   study the shock front,
which requires X-ray diagnostics of high  temporal and spatial
resolution and is thus  very challenging. An
important result of  our  study is the reduction with radiative
losses in   the time  needed  to reach   the  stationary  limit. Although  some experiments  have been
close  to reaching  this limit  \citep{lpb},  this has  not been
studied until now experimentally.

Shock  velocities of 60  - 80 km/s   are accessible with kJ
class  lasers.  Using more  powerful  laser-energy installations such as
the NIF \citep{NIF} at Livermore (USA) or the LMJ \citep{LMJ3, LMJ1, LMJ2}
at
Bordeaux (France) will allow  us to explore the  effects of the radiation
pressure on the structure of the shock wave.  Using xenon at densities
of the   order of 5x10$^{-4}  $   g/cm$^3$, such effects   start to be
visible for shock velocities of the order of 200 km/s \citep{michaut}.

Another interesting topic that could be addressed is the eventual
development of     radiative      instabilities.   These   ``thermal''
instabilities  have  been  studied theoretically  and  numerically for
optically thin radiative shocks, in  the  context of, for example,  accretion  onto
white dwarfs,  and  colliding   radiative flows   \citep{falle,
chevalier,  bertschinger, ryu,   walder,  laming,  mignone}.   Such an
instability was observed experimentally  in 3D blast-wave development
over  a few hundreds  of  nanoseconds \citep{grun}, although  other  experiments
failed to reproduce  it on a  shorter timescale \citep{edwards}. The
development of these  instabilities in an optically-thicker  radiative
shock (with a  radiative precursor)  has,  to our  knowledge, not been
studied theoretically.  The multidimensional effects, by deforming the
front and allowing a transition towards  an optically thin case, could
be favorable to the development of such  instabilities, and this will be the
subject of future studies.

These laboratory studies allow us to test the adequacy of modern codes in
dealing with  astrophysical, hypersonic  radiative  flows.  The HERACLES 3D
 radiation hydrodynamics code  is  well adapted  to the  study of these
anisotropic radiative flows.   It allows us to test  together the cases of both
optically thick and optically thin regimes.   It has been already used
in  modeling the propagation of  astrophysical radiative jets in
the   ISM  \citep{sf2a_2006_pcmi,gonzalez}.  Future improvements
will  include the inclusions of  multigroup radiative-transport (which
may influence the fine structure of the shock precursor), the decoupling of
ionic and electronic temperatures (which  affect the shock front), and
non-LTE effects, which  play a role in  the optically thin regions  of
the shock structure.

\begin{acknowledgements}
The authors are indebted to M.~Busquet and to the anonymous referee for valuable comments and suggestions leading to an improvement of this manuscript.
This work was  carried out under   the EU funded  RTN JETSET (contract
MRTN-CT-2004 005592).  The  authors acknowledge the financial  support
of CNRS program PNPS and ANR grant SiNERGHy (ANR-06-CIS6-009-01) and would like to thanks the CEA computing center
(CCRT) where all the simulations were done.
M.~G. acknowledges the financial support provided by the European Commission TUIXS project and the French Ministry of Foreign Affairs through the Lavoisier grant.
\end{acknowledgements}

\bibliographystyle{aa}
\bibliography{9136} 

\end{document}